# Energy gaps, topological insulator state and zero-field quantum Hall effect in graphene by strain engineering

F. Guinea[1], M. I. Katsnelson[2] & A. K. Geim[3]

[1]*Instituto de Ciencia de Materiales de Madrid (CSIC), Sor Juana Inés de la Cruz 3, Madrid 28049, Spain*
[2]*Institute for Molecules and Materials, Radboud University Nijmegen, Nijmegen, The Netherlands*
[3]*Manchester Centre for Mesoscience and Nanotechnology, University of Manchester, Manchester, UK*

**Among many remarkable qualities of graphene, its electronic properties attract particular interest due to a massless chiral character of charge carriers, which leads to such unusual phenomena as metallic conductivity in the limit of no carriers and the half-integer quantum Hall effect (QHE) observable even at room temperature [1-3]. Because graphene is only one atom thick, it is also amenable to external influences including mechanical deformation. The latter offers a tempting prospect of controlling graphene's properties by strain and, recently, several reports have examined graphene under uniaxial deformation [4-8]. Although the strain can induce additional Raman features [7,8], no significant changes in graphene's band structure have been either observed or expected for realistic strains of ~10% [9-11]. Here we show that a designed strain aligned along three main crystallographic directions induces strong gauge fields [12-14] that effectively act as a uniform magnetic field exceeding 10 T. For a finite doping, the quantizing field results in an insulating bulk and a pair of counter-circulating edge states, similar to the case of a topological insulator [15-20]. We suggest realistic ways of creating this quantum state and observing the pseudo-magnetic QHE. We also show that strained superlattices can be used to open significant energy gaps in graphene's electronic spectrum.**

If a mechanical strain Δ varies smoothly on the scale of interatomic distances, it does not break the sublattice symmetry but rather deforms the Brillouin zone in such a way that the Dirac cones located in graphene at points *K* and *K'* are shifted in the opposite directions [2]. This is reminiscent of the effect induced on charge carriers by magnetic field *B* applied perpendicular to the graphene plane [2,12-14]. The strain-induced, pseudo-magnetic field $B_S$ or, more generally, gauge field vector potential **A** have opposite signs for graphene's two valleys *K* and *K'*, which means that elastic deformations, unlike magnetic field, do not violate the time-reversal symmetry of a crystal as a whole [12-14,21,22].

Based on this analogy between strain and magnetic field, we ask the following question: Is it possible to create such a distribution of strain that it results in a strong uniform pseudo-field $B_S$ and, accordingly, leads to a "pseudo-QHE" observable in zero *B*? The previous attempts to engineer energy gaps by applying strain [5-7] seem to suggest a negative answer. Indeed, the hexagonal symmetry of the graphene lattice generally implies a highly anisotropic distribution of $B_S$ [21,22]. Therefore, the strain is expected to contribute primarily in the phenomena that do not average out in a random magnetic field such as weak localization [13,14]. Furthermore, a strong gauge field necessitates the opening of energy gaps due to Landau quantization, $\delta E \approx 400\text{K} \cdot \sqrt{B}$ (>0.1 eV for $B_S$ =10T) whereas no gaps were theoretically found for uniaxial strain as large as ≈25% [4]. The only way to induce significant gaps, which was known so far, is to spatially confine carriers ($\delta E \approx 0.1$ eV requires 10 nm wide ribbons) [1,2]. Contrary to these expectations, we have found that by applying stresses with triangular symmetry, it is possible to generate a uniform quantizing $B_S$ equivalent to tens of Tesla so that the corresponding gaps exceed 0.1 eV and are observable at room temperature.

A two-dimensional strain field $u_{ij}(x,y)$ leads to a gauge field [23,24]

$$\mathbf{A} = \frac{\beta}{a}\begin{pmatrix} u_{xx} - u_{yy} \\ -2u_{xy} \end{pmatrix} \qquad (1)$$

where *a* is the lattice constant, $\beta = -\partial \ln t / \partial \ln a \approx 2$, *t* the nearest-neighbour hopping parameter, and the *x*-axis is chosen along a zigzag direction of the graphene lattice. In the following, we consider valley *K*, unless stated otherwise. One can immediately see that $B_S$ can only be created by non-uniform shear strain. Indeed, for dilation (isotropic strain), equation (1) leads to $\mathbf{A} = 0$ and, for the uniform strain previously considered in refs. [4-6], to $\mathbf{A} = const$ which also yields zero $B_S$.

Using polar coordinates $(r, \theta)$, equation (1) can be rewritten as

$$A_r = \frac{\beta}{a}\left[\left(\frac{\partial u_r}{\partial r} - \frac{u_r}{r} - \frac{1}{r}\frac{\partial u_\theta}{\partial \theta}\right)\cos 3\theta + \left(-\frac{1}{r}\frac{\partial u_r}{\partial \theta} + \frac{u_\theta}{r} - \frac{\partial u_\theta}{\partial r}\right)\sin 3\theta\right],$$
$$A_\theta = \frac{\beta}{a}\left[\left(-\frac{\partial u_\theta}{\partial r} + \frac{u_\theta}{r} - \frac{1}{r}\frac{\partial u_r}{\partial \theta}\right)\cos 3\theta + \left(\frac{1}{r}\frac{\partial u_\theta}{\partial \theta} + \frac{u_r}{r} - \frac{\partial u_r}{\partial r}\right)\sin 3\theta\right] \quad (2)$$

which yields the pseudo-magnetic field

$$B_S = \frac{\partial A_y}{\partial x} - \frac{\partial A_x}{\partial y} = \frac{1}{r}\frac{\partial A_r}{\partial \theta} - \frac{\partial A_\theta}{\partial r} - \frac{A_\theta}{r} \quad (3).$$

In the radial representation, it is easy to show that uniform $B_S$ is achieved for the following displacements:

$$u_r = cr^2 \sin 3\theta, \quad u_\theta = cr^2 \cos 3\theta \quad (4)$$

where $c$ is a constant. The strain described by (4) and its crystallographic alignment are shown in Figures 1a and 1b, respectively. This yields uniform $B_S = 8\beta c/a$ (given in units $\hbar/e \equiv 1$). For a disk of diameter $D$, which experiences a maximum strain $\Delta_m$ at its perimeter, we find $c = \Delta_m/D$. Assuming achievable $\Delta_m$ = 10% and $D$ = 100 nm, we find $B_S \approx 40$T, the effective magnetic length $l_B = \sqrt{\frac{aD}{8\beta\Delta_m}} \approx 4$ nm and the largest Landau gap of $\approx 0.25$ eV. Note that distortions (4) are purely shear and do not result in any changes in the area of a unit cell, which means that there is no effective electrostatic potential generated by such strain [23].

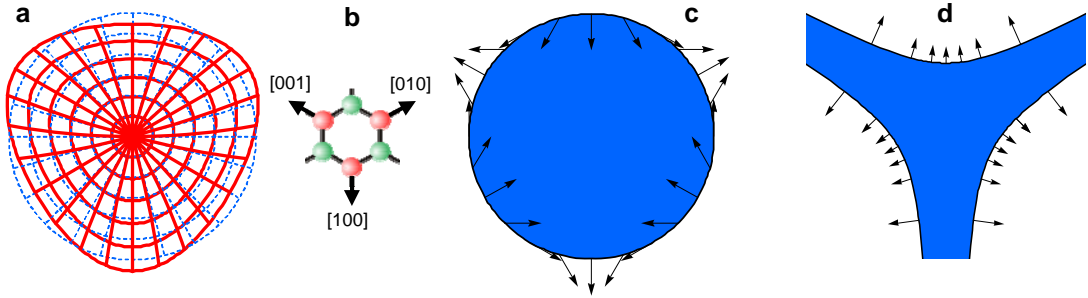

**Figure 1**. Designed strain can generate a strictly uniform pseudo-magnetic field in graphene. (**a**) Distortion of a graphene disc which is required to generate uniform $B_S$. The original shape is shown in blue. (**b**) Orientation of the graphene crystal lattice with respect to the strain. Graphene is stretched or compressed along three equivalent crystallographic directions <100>. Two graphene sublattices are shown in red and green. (**c**) Distribution of the forces applied at disk's perimeter (arrows), which would create the strain required in (**a**). The uniform colour inside the disk indicates strictly uniform pseudo-magnetic field. (**d**) The shown shape allows uniform $B_S$ to be generated only by normal forces applied at the sample's perimeter. The length of arrows indicates the required local stress.

The lattice distortions in Fig. 1a can be induced by in-plane forces $\mathbf{F}$ applied only at the perimeter and, for the case of a disc, they are given simply by

$$F_x(\theta) = 4c\mu\sin(2\theta), \qquad F_y(\theta) = 4c\mu\cos(2\theta) \tag{5}$$

where $\mu$ is the shear modulus. Fig. 1c shows the required force pattern. It is difficult to create such strain experimentally because this involves tangential forces and both stretching and compression. To this end, we have solved an inverse problem to find out whether uniform $B_S$ can be generated by normal forces only. There exists a unique solution for the shape of a graphene sample, which allows this and is plotted in Fig. 1d (see Supplementary Information, part I).

A strong pseudo-magnetic field should lead to Landau quantization and a QHE-like state. The latter is different from the standard QHE because $B_S$ has opposite signs for charge carriers in valleys $K$ and $K'$ and, therefore, generates edges states that circulate in opposite directions. The co-existence of gaps in the bulk and counter-propagating states at the boundaries without breaking the time-reversal symmetry is reminiscent of the topological insulators [15-20] and, in particular, the quantum valley Hall effect in "gapped graphene" [20] and the quantum spin Hall effect induced by strain [16]. The latter theory has exploited the influence of three-dimensional strain on spin-orbit coupling in semiconductor heterostructures, which can lead to quasi-Landau quantization with opposite $B_S$ acting on two spins rather than valleys. Weak spin-orbit coupling allows only tiny Landau gaps <1 μeV [16] which, to be observable, would require temperatures below 10 mK and carrier mobilities higher than $10^7$ cm$^2$/Vs. Our approach exploits the unique strength of pseudospin-orbit coupling in graphene, which leads to $\delta E > 0.1$ eV and makes the strain-induced Landau levels realistically observable.

The two cases described above prove that by using strain it is possible to generate a strong uniform $B_S$ and observe the pseudo-QHE. They also provide the general concept that if the strain is applied along all three <100> crystallographic directions to match graphene's symmetry, this prevents the generated fields from changing sign. Experimentally, it is a difficult task to generate such a complex distribution of forces as shown in Fig. 1. Below we develop the above concept further and show that the pseudo-QHE can be observed in geometries that are easier to realize, even though they do not provide a perfectly uniform $B_S$.

Let us consider a regular hexagon with side length $L$ and normal stresses applied evenly at its three non-adjacent sides and along <100> axes (Fig. 2a). Our numerical solution for this elasticity problem shows that $B_S$ has a predominant direction (positive for $K$ and negative for $K'$) and is rather uniform close to the hexagon's centre. Assuming $L = 100$ nm and $\Delta_m = 10\%$, we find for Fig. 2a that $B_S$ varies between ±22 T but is ≈20 T over most of the hexagon's central area. For other $L$ and $\Delta$, one can rescale the plotted values of $B_S$ by using expression $B_S \propto \Delta_m / L$. We have also examined other geometries (Supplementary Information) and always found a nearly uniform distribution of $B_S$ near samples' centre.

To verify that the non-uniform $B_S$ in Fig. 2a leads to well-defined Landau quantization, we have calculated the resulting density of states $D(E)$. This analysis necessitates tight-binding calculations that were carried out for a graphene hexagon with zigzag edges (see Supplementary Information). Our computational power has limited the size of the studied hexagons to $L ≈ 30$ nm. Fig. 2b plots our results for $\Delta_m = 1\%$ ($B_S ≈ 7$T at the hexagon centre) and compares them for the case of the same hexagon in $B = 0$ and 10T but without strain. In the absence of strain or $B$, the peak at zero $E$ is due to the states localized at zigzag edges [2]. One can also see that both non-uniform $B_S$ and uniform $B$ generate Landau levels and the quality of the induced quantization is rather similar. The zero-$E$ peak in Fig. 2b is dominated by zigzag states in all three cases whereas the other levels are slightly broadened by non-uniform $B_S$. The influence of the inhomogeneity in $B_S$ on the zero level should, in general, be minimal because magnetic field inhomogeneity does not lead to broadening of this level [25]. We emphasize that the presence of a significant density of states between pseudo-Landau levels in Fig. 2b is mostly due to small $L$ used in our numerical calculations. For micrometer hexagons, the corresponding gaps (even when averaged over the whole sample) should be well resolved if only slightly smeared by non-uniform $B_S$.

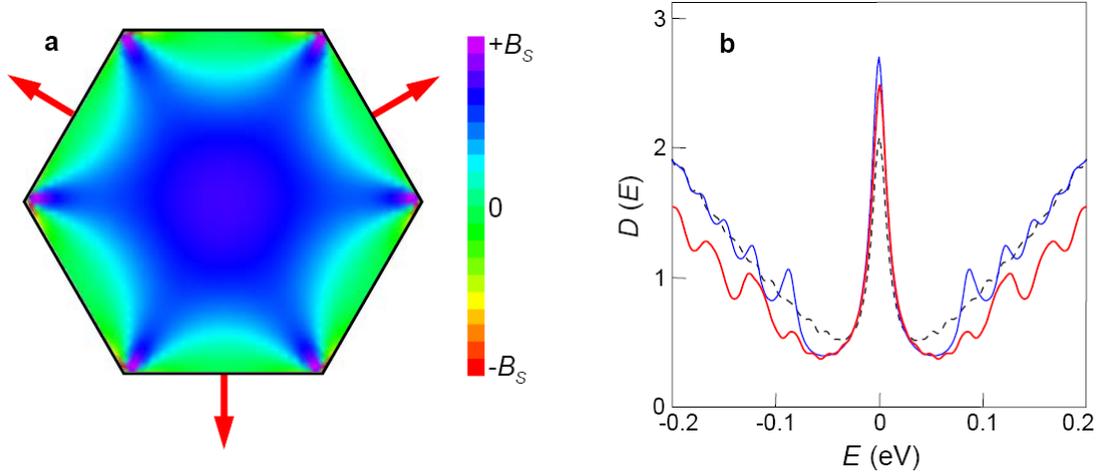

**Figure 2**. Stretching graphene along <100> axes generates a pseudo-magnetic field that is rather uniform at the centre. (**a**) Distribution of $B_S$ for a regular hexagon stretched by its three sides oriented perpendicular to <100>. Other examples are given in Supplementary Information. (**b**) Normalized density of states for the hexagon in (**a**) with $L$ = 30 nm and $\Delta_m$ = 1%. The black curve is for the case of no strain and no magnetic field. The peak at zero $E$ is due to states at zigzag edges. The blue curve shows the Landau quantization induced by magnetic field $B$ = 10 T. The pseudo-magnetic field with $B_S \approx 7$ T near the hexagon's centre induces the quantization shown by the red curve. A comparison between the curves shows that the finite density of states between the pseudo-Landau levels is due to the small sample size in the tight-binding calculations.

In order to create the required strain experimentally, one can generally think of exploiting the difference in thermal expansion of graphene and a substrate [11] and apply temperature gradients along <100> axes. For the case of quasi-uniform $B_S$, there are many more options available, including the use of suspended samples and profiled substrates. For example, a graphene hexagon can be suspended by three metallic contacts attached to its sides, similar to the technique used to study suspended graphene [26,27], and the strain can then be controlled by gate voltage. Alternatively, a quasi-uniform $B_S$ can be created by depositing graphene over triangular trenches (Supplementary Information).

To probe the pseudo-Landau quantization, one can employ optical techniques, for example, Raman spectroscopy that should reveal extra resonances induced by $B_S$ [28]. One can also use transport measurements in both standard and Corbino-disk geometries. In the former case, the counter-propagating edge states imply that contributions from two valleys cancel each other and no Hall signal is generated ($\rho_{xy}$=0) [15-20]. At the same time, the edge transport can lead to longitudinal resistivity $\rho_{xx} = h/4e^2N$ where $N$ is the number of spin-degenerate Landau level at the Fermi energy. This non-zero quantized $\rho_{xx}$ has the same origin as in so-called dissipative QHE where two edge states with opposite spins propagate in opposite directions [29]. In spin-based topological insulators, the edge transport is protected by slow spin flip rates [15,16,29]. In our case, atomic-scale disorder at the edges is likely to mix the counter-circulating states on a submicron scale (Supplementary Information). Therefore, instead of quantization in $\rho_{xx}$ we may expect highly-resistive metallic edge states, similar to the case discussed in ref. [29]. The suppression of the edge-state ballistic transport does not affect the pseudo-Landau quantization in graphene's interior, where intervalley scattering is very weak [13,30] and should not case any extra level broadening. Highly-resistive edges should in fact make it easier to probe pseudo-Landau gaps in the bulk. In the Corbino geometry, the edge-state mixing is irrelevant, and we expect two-probe $\rho_{xx}$ to be a periodic function of gate voltage and show an insulating behaviour between pseudo-Landau levels. Furthermore, the outer contact can be used to cover perimeter regions with non-uniform $B_S$ (such as in Fig. 2) which should improve the quality of quantization.

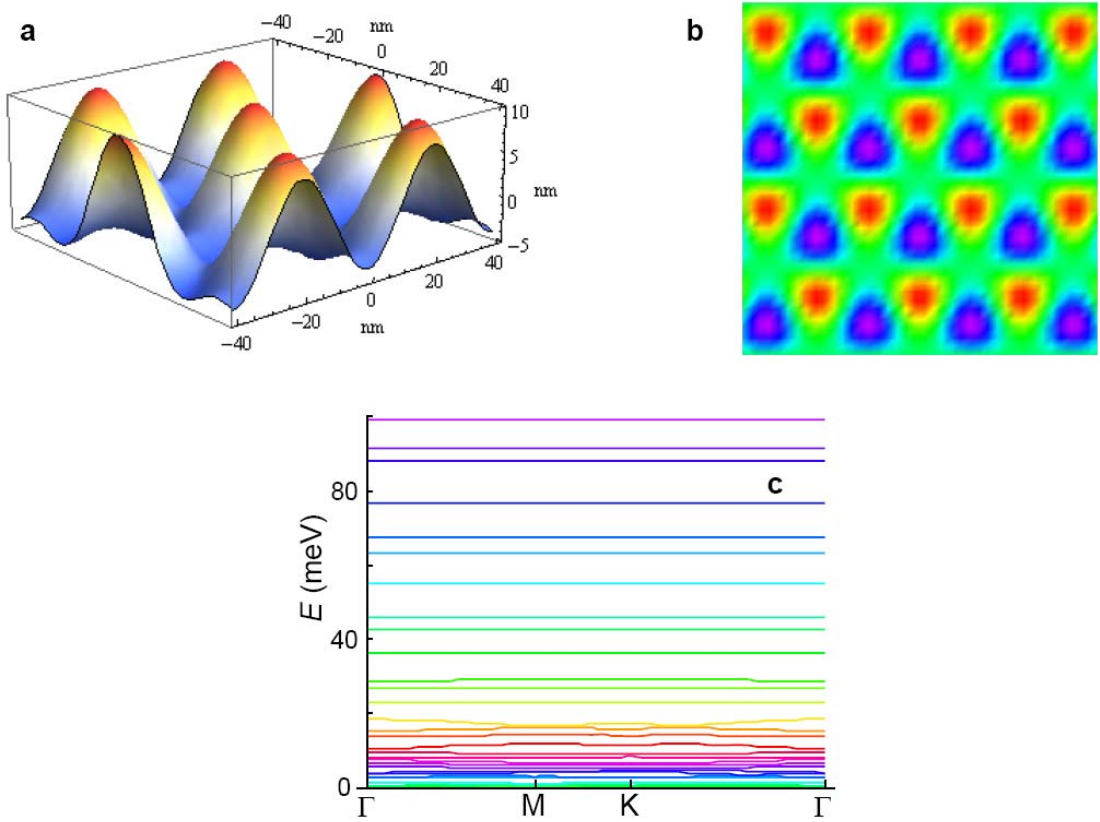

**Figure 3**. Energy gaps can be opened in strained graphene superlattices. (**a**) Strain with triangular symmetry can be created by depositing graphene on profiled surfaces. The corrugations shown in (**a**) result in the distribution of pseudo-magnetic field plotted in (**b**). $B_S$ varies between ±0.5T (red to violet) with the periodicity twice shorter than in (**a**). (**c**) Low energy bands induced by the periodic strain. The bands are symmetric with respect to zero $E$.

Finally, we point out that the developed concept can be employed to create gaps in bulk graphene. Imagine a macroscopic graphene sheet deposited on top of a corrugated surface with a triangular landscape (Fig. 3a). In the following calculations, we have fixed the graphene sheet at the landscape's extrema and allowed the resulting in-plane displacements to relax [21,22] (at the nanoscale, graphene should then be kept in place by van der Waals forces). The resulting pseudo-magnetic superlattice is plotted in Fig. 3b whereas Fig. 3c shows the resulting energy spectrum. Close to zero $E$, there is a continuous band of electronic states, in agreement with the fact that zero level is insensitive to field's inhomogeneity [25]. At higher $E$, there are multiple gaps with $\delta E$ >100 K. The relatively small gaps are due to the weak shear strain induced in this geometry ($\Delta_m$ <0.1%). By improving the design of strained superlattices, it must be possible to achieve much larger gaps. We believe that the suggested strategies to observe the pseudo-Landau gaps and QHE are completely attainable and will be realized sooner rather than later.

*Acknowledgements*
This work was supported by the EPSRC (UK), the Royal Society, Office of Naval Research and Air Force Office of Scientific Research.
*Author contributions*
All authors contributed extensively to the work presented in this paper.

**SUPPLEMENTARY INFORMATION:**
**Energy gaps and zero-field quantum Hall effect in graphene by strain engineering**
F. Guinea, M. I. Katsnelson, A. K. Geim

**I**. Let us explain first how the two dimensional elasticity theory was used to find the shape shown in Fig. 1d of the main text.
The distribution of stress $\sigma$ in a two dimensional case is known to be independent of material's Poisson ratio [S1]. The distortions that determine the gauge field are proportional to $\sigma$ and given by

$$u_{xx} - u_{yy} = \frac{\sigma_{xx} - \sigma_{yy}}{2\mu}, \quad u_{xy} = \frac{\sigma_{xy}}{2\mu}, \tag{S1}$$

where $\mu$ is the shear modulus. Its value for graphene is $\approx 10$ eV/Å$^2$ (see ref. [S2]).

According to equations (1) and (3) of the main text and after choosing a coordinate system such that the $x$-axis coincides with one of the zigzag directions, we find that the uniformity of pseudo-magnetic field requires the following stress distribution

$$\sigma_{xx} = -\sigma_{yy} = Cy$$
$$\sigma_{xy} = Cx \tag{S2}$$

where $C$ is a constant.
Let us find a boundary at which this stress distribution creates only normal forces. We will use polar coordinates in which the boundary is described as $r = r(\theta)$ with the normal vector $(n_x(\theta), n_y(\theta))$. Then, the condition that the forces are strictly normal to the boundary reads

$$\sigma_{xx}n_x + \sigma_{xy}n_y = f(\theta)n_x$$
$$\sigma_{xy}n_x + \sigma_{yy}n_y = f(\theta)n_y \tag{S3}$$

This means that the stress tensor at this boundary has the following structure

$$\sigma_{ij} = 2f(\theta)\left(n_i n_j - \frac{1}{2}\delta_{ij}\right) \tag{S4}$$

This equation has a solution if
$$(\sigma_{xx} - f)(\sigma_{yy} - f) - \sigma_{xy}^2 = 0$$
or, substituting our sigma's, it can be re-written as $f(\theta) = \pm Cr(\theta)$.

Then, $\dfrac{n_x}{n_y} = \dfrac{\pm r + y}{x} = \dfrac{\pm 1 + \sin\theta}{\cos\theta}$ and, at the same time, $\dfrac{n_x}{n_y} = -\dfrac{dy}{dx}$. Coming back to the polar coordinates, we find the following equation for the required shape

$$\frac{d\ln r}{d\theta} = \frac{\sin\theta \mp \cos 2\theta}{\cos\theta \pm \sin 2\theta}. \tag{S5}$$

Its solution

$$r(\theta) = \frac{const}{[(\cos\theta/2 \mp \sin\theta/2)(\pm 1 + 2\sin\theta)]^{2/3}} \tag{S6}$$

is plotted in Fig. 1d. The required distribution of normal forces is given by $F(\theta) \propto \pm r(\theta)$ where sign $\pm$ indicates that uniform $B_S$ can be achieved by both compression and stretching.

**II**. To calculate the distribution of pseudo-magnetic field $B_S$, which is shown in Fig. 2a, we have solved the corresponding elasticity problems numerically.
The stress distribution $\sigma_{ij}(\vec{r})$ induced inside the hexagon in Fig. 2a is caused by forces $F$ applied at the perimeter in the direction normal to the edge. They are either zero or constant for different hexagon sides.

The stress distribution can be written as $\sigma_{ij}(\vec{r}) = F s_{ij}(\vec{r}/L)$ where $L$ is the length of the hexagon side. Using equation (S1) we can then find the strain distribution in graphene. To calculate $s_{ij}(\vec{u})$, we have used a triangular mesh with 35 nodes along each side of the hexagon and the central forces acting between the nearest neighbour nodes. This has yielded $B_S$ plotted in Fig. 2a. Another example of pseudo-magnetic field in a strained graphene sample is shown below (Fig. S1).

The electronic states inside the hexagon in Fig. 2b were calculated by using a honeycomb lattice with the same periodicity as the triangular lattice used to obtain the induced strain in Fig. 2a. In order to make these calculations applicable to hexagons with sides larger than 35 graphene lattice constants $a$, we have exploited the scaling properties of the low energy eigenstates of the Dirac equation in effective gauge field $\mathbf{A}$. The eigenenergies $\varepsilon_i$ in a lattice of length $L = N \times a$ described by hopping parameters $t_{ij} = \bar{t} + \Delta t_{ij}$ are related to those in a lattice with $L' = N' \times a$ and hopping parameters $t'_{ij} = \bar{t}' + \Delta t'_{ij}$ by $\varepsilon'_i = \varepsilon_i \times N/N'$ provided that $t'_{ij} = \bar{t} \times N/N' + \Delta t_{ij} \times N'/N$. This scaling relationship ensures that the Fermi velocity and the flux distribution remain the same within the two lattices. The equivalence between the two systems is valid only for $\varepsilon'_i \ll \bar{t}'$ and $\Delta t'_{ij} \ll \bar{t}'$. These constraints have limited the maximum size that could be studied in our work to about $L' \approx 10 \times N \times a$ and maximum strain to $\Delta \approx \beta \Delta t'_{ij}/\bar{t}' \approx 0.01$.

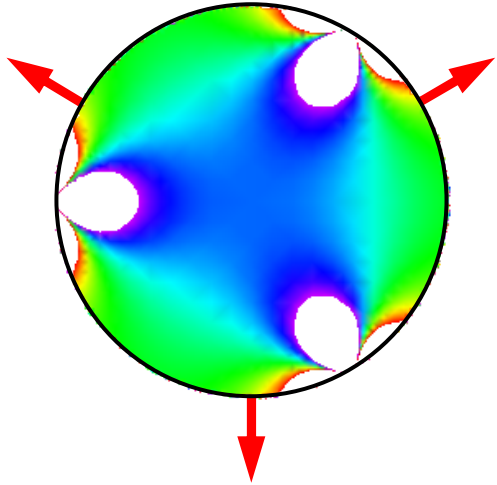

Figure S1. Distribution of $B_S$ inside a graphene disk stretched along the <100> axes indicated by the red arrows. The colour scheme is the same as in Fig. 2a. The white (blank) areas correspond to divergences in the induced strain and $B_S$ because the forces are applied locally, at three points along the perimeter. In the real situation, these divergences would correspond to high values of $B_S$, which depend on detailed distribution of stress at the sample boundary.

**III**. An interesting alternative to generate the pseudo-magnetic field is to apply a constant pressure $P$ to graphene suspended over an aperture with designed geometry. In this case, $B_S$ can be calculated as follows. The distribution of in-plane $\mathbf{u}(x,y)$ and out-of-plane $h(x,y)$ deformations can be found numerically by minimization of the free energy [S3]

$$F = \frac{\kappa}{2} \int dxdy (\nabla^2 h)^2 + \frac{1}{2} \int dxdy (2\mu u_{ij}^2 + \lambda u_{ii}^2) - P \int dxdy\, h \qquad (S6),$$

where $\kappa, \lambda, \mu$ are the bending rigidity and the two in-plane elastic constants, respectively, and

$$u_{ij} = \frac{1}{2}\left( \frac{\partial u_i}{\partial x_j} + \frac{\partial u_j}{\partial x_i} + \frac{\partial h}{\partial x_i} \frac{\partial h}{\partial x_j} \right) \qquad (S7)$$

is the deformation tensor. If out-of-plane displacements are larger than the length scale given by $\sqrt{\kappa/\mu}$, which is of the order of the interatomic distance $a$, we can generally neglect the first term in equation (S6) [S1].

Assuming that a graphene membrane covers an aperture of a characteristic size $L$, the out-of-plane deformations can be represented as

$$h(x,y) = L\left(\frac{PL}{\mu}\right)^{1/3} \overline{H}\left(\frac{x}{L},\frac{y}{L},\nu\right) \tag{S8}$$

where $\overline{H}$ is the dimensionless function that depends on the membrane's geometry, and $\nu$ the Poisson ratio. Following refs. [S4-S6], we find that equation (S8) leads to the pseudo-magnetic field

$$B_S(x,y) = \beta\frac{\Phi_0}{aL}\left(\frac{PL}{\mu}\right)^{2/3} \overline{B}\left(\frac{x}{L},\frac{y}{L},\nu\right) \tag{S9}$$

where $\Phi_0 = \pi\hbar c/|e|$ is the flux quantum and, to avoid a bulky expression, we introduce another dimensionless function $\overline{B}$ that relates to $\overline{H}$.

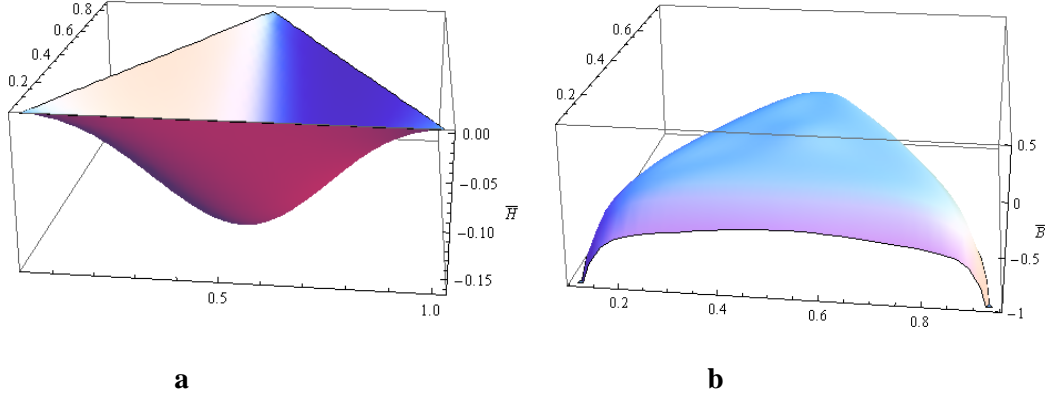

**a**  **b**

Figure S2. Deformation and pseudo-magnetic field induced in graphene suspended over a triangular aperture. (**a**) Height profile $\overline{H}$ according to equation (S8). (**b**) Pseudo-magnetic field $\overline{B}$ for the deformation shown in **a**. The spatial scale for both plots is given in units $L$.

We have calculated $\overline{H}(x,y)$ and $\overline{B}(x,y)$ numerically for the case of an equilateral triangular aperture with the sides normal to the <100> axes of the graphene membrane. Our results in Figure S2 demonstrate that the pseudo-magnetic field is nearly uniform inside the hole, yielding $\overline{B}$ of about 0.3. Assuming $L = 100$ nm and $P = 100$ atm, we find $B_S \approx 4$ T. The induced strain can be estimated as $(PL/\mu)^{2/3}$ and, at this pressure, reaches only a few percent, that is, even higher $B_S$ can be achieved at higher $P$. Note that the pressure on a suspended membrane can also be induced by electric field [S6]. In terms of charge density $n$, this pressure is given by $P = 2\pi(ne)^2$, and $P = 100$ atm corresponds to $n \approx 8\times10^{12}$ cm$^{-2}$.

**IV**. To calculate $\overline{H}$ and $\overline{B}$ in Figure S2, we have employed the elastic lattice model described previously in section II. We have taken a triangular lattice with nodes $\{m,n\}$ placed at positions $\vec{r}_{mn} = (m+n/2, \sqrt{3}n/2)$ where $m,n$ are integers. For each node, we assume displacement $\vec{u}_{mn}$. The elastic energy of the lattice is the sum of energies over all the springs connecting nearest neighbour nodes, which are given by $\varepsilon_{kl;mn} = \frac{K}{2}\left(|\vec{r}_{kl} + \vec{u}_{kl} - \vec{r}_{mn} - \vec{u}_{mn}| - 1\right)^2$ where $K$ is related to the elastic constants as $\lambda = \mu = \frac{\sqrt{3}}{4}K$ [S7]. The pinning of the graphene sheet to the scaffold at the perimeter nodes $\{o,p\}$ is described by adding term $\varepsilon_{pin,op} = \frac{K_{pin}}{2}|\vec{u}_{op}|^2$, and the constant pressure (force) is applied in $z$ direction. The calculations were done for a membrane with 820 nodes ($L = 40$), $K = 1$ and $K_{pin} = 100$. The numerical results show that $\max[\overline{H}(x,y)] \approx 0.15 \pm 0.03$ and the field varies slowly inside the hole, with an average $\langle|\overline{B}(x,y)|\rangle \approx 0.3 \pm 0.03$. Note that the triangular-lattice model implies $\lambda = \mu$ and, therefore, $\nu = 1/3$. In graphene, the Poisson ratio is ~0.15 [S2] but we do not expect that this discrepancy would result in significantly different values of $\overline{H}$, $\overline{B}$ and $B_S$.

**V.** In the following, we estimate a mixing rate for counter-propagating valley-polarized edge states.
In the case of a singly-charged impurity, the matrix element that determines the backscattering is given by $\langle k_+|e^2/|\vec{r}||k_-\rangle \approx e^2 a$ (the wavefunctions are assumed to be extended along the edge channel). The density of states is $D(E) \approx v_F \ell_B$, where $\ell_B$ is the width of the edge channel and $v_F$ the Fermi velocity. Then, the reflection coefficient $R$ into another valley can be estimated in the Born approximation as $R \approx (e^2/v_F)^2 (a/\ell_B)^2$. The number of impurities per unit length is $n_{1D} \approx n\ell_B$ where $n$ is the impurity concentration. The transmission decays with the channel length $l$ as $T \approx \exp(-Rn_{1D}l) \approx \exp[-(e^2/v_F)^2 na^2 l/\ell_B]$. For graphene, $e^2/v_F \approx 1$, and the mean free path in our one-channel one-dimensional problem becomes of the same order of magnitude as the localization length $\xi \approx \ell_B/(na^2)$. For typical $B_s = 10$T ($\ell_B \approx 8$ nm) and $n \approx 10^{11}$ cm$^{-2}$, we find $\xi \approx 10^2 - 10^3$ μm.
In the case of edge roughness, it can be modeled by a succession of vacancies. We approximate each vacancy by a local potential with a strength comparable to the bandwidth of the π electrons, $v_F/a$. Using the Born approximation, we find that the scattering matrix element is $\langle k_+|v_F/a|k_-\rangle \approx v_F a$ and $R \approx a^2/\ell_B^2$, so that the mean free path is now $\xi \approx (\ell_B^2/a^2)/n_{1D}$. The density of defects (vacancies) $n_{1D}$ is expected to be strongly sample dependent. In the worst case scenario of atomic-scale edge roughness, $n_{1D} \approx 10$ per nm and we still find $\xi \approx 10^2 - 10^3$ nm. The reason for this is that the scattering is suppressed by the smallness of $a/\ell_B$. Note that edge roughness does not influence the strain-induced Landau quantization in graphene's interior.